\documentclass[secnumarabic,amssymb, nobibnotes, aps, pre,notitlepage]{revtex4-1}
\usepackage{tikz,amsmath}
\usepackage{epstopdf}

\begin{document}

\title{The Many Faces of Far-from-equilibrium Thermodynamics: Deterministic Chaos, Randomness or Emergent Order?}

\author{Atanu Chatterjee\email[Email:]{achatterjee3@wpi.edu} and Germano Iannacchione\email[Email:]{gsiannac@wpi.edu}}
\affiliation{Department of Physics and the Order-Disorder Phenomena Lab\\ Worcester Polytechnic Institute, Worcester, MA, USA, 01609}
\date{\today}

\begin{abstract}
\noindent Far-from-equilibrium systems are ubiquitous in nature. They are also rich in terms of diversity and complexity. Therefore, it is an intellectual challenge to be able to understand the physics of far-from-equilibrium phenomena. In this paper we revisit a standard tabletop experiment, the Rayleigh-B{\'e}nard convection, to explore some fundamental questions and present a new perspective from a first-principles point of view. How non-equilibrium fluctuations differ from equilibrium fluctuations, how emergence of order out-of-equilibrium breaks symmetries in the system, or how free-energy of a system gets locally bifurcated to operate a Carnot-like engine to maintain order? The exploration and investigation of these non-trivial questions are the focus of this paper.
\end{abstract}

\maketitle

\section{Introduction}
\begin{figure}[b]
\centering
\includegraphics[scale=0.6]{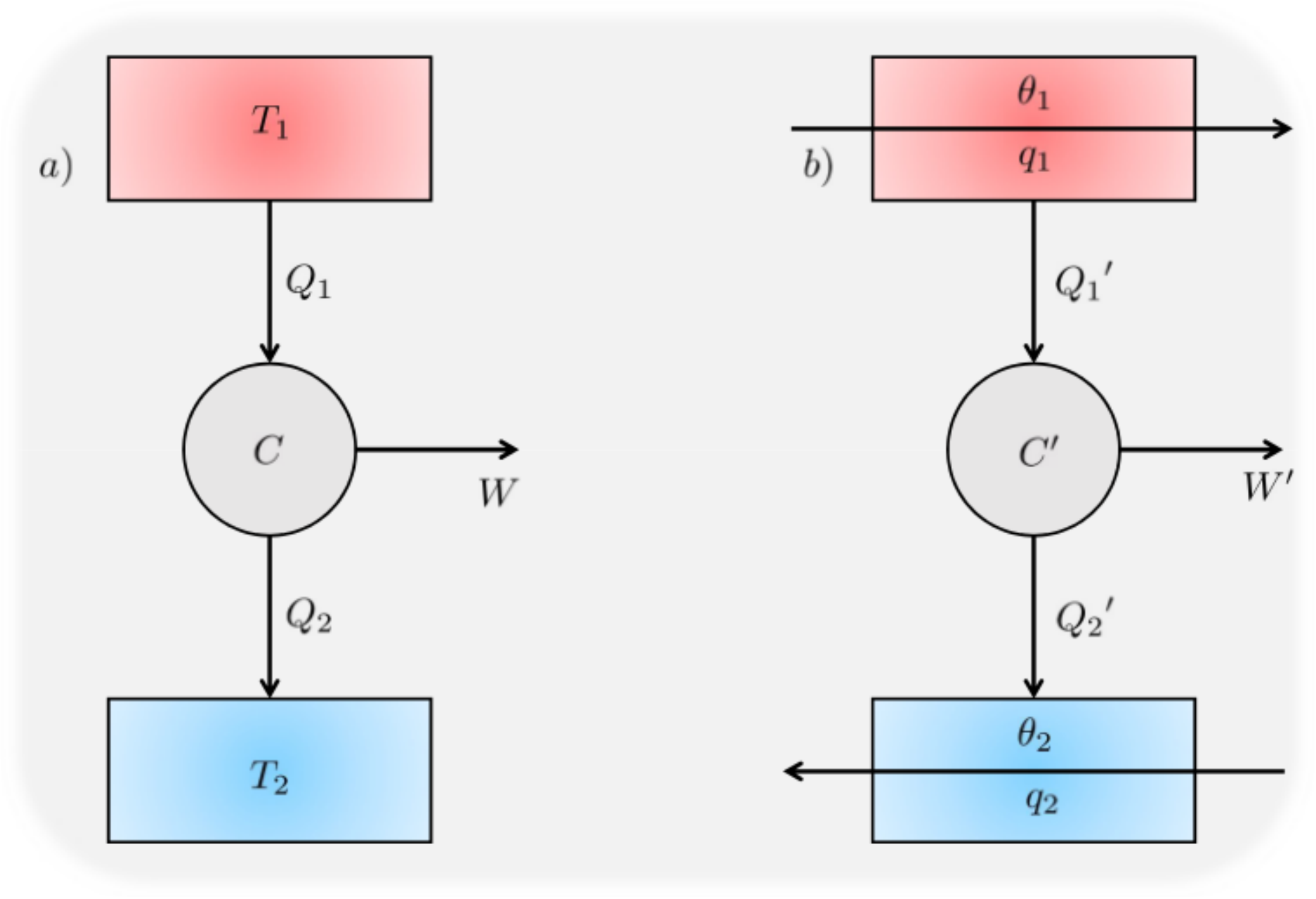}
\caption{a) Figure shows a theoretical Carnot engine, $C$, operating between the thermal reservoirs $T_1$ and $T_2$ ($T_1>T_2$). It derives heat $Q_1$ from the reservoir kept at $T_1$, rejects heat, $Q_2$ into the reservoir kept at $T_2$ while performing work, $W$. b) Figure shows a practical Carnot engine, $C^\prime$, operating between two thermal reservoirs $\theta_1$ and $\theta_2$ kept at a steady-state by the constant heat influxes, $q_1$ and $q_2$. It derives heat $Q_1^\prime$ from the reservoir kept at $\theta_1$, rejects heat, $Q_2^\prime$ into the reservoir kept at $\theta_2$ while performing less work, $W^\prime$ ($W^\prime<W$).}
\end{figure}
\noindent A system, isolated from its surrounding will continue to be in a state of equilibrium unless driven by an external steady flow of energy. Statistically, a state of equilibrium implies a state of randomness, and randomness implies symmetry. Therefore, from a microscopic sense all states in a statistical ensemble are equally likely, and the system explores all possibilities before collapsing into a single point in the phase space, characterized by the macroscopic thermodynamic variables like, pressure, temperature and volume. The physics of such equilibrium systems forms the basis of classical thermodynamics and statistical mechanics~\cite{clausius1854veranderte,planck2013treatise,gibbs1906scientific,martyushev2006maximum,lieb1998guide}. However, if we look around ourselves we are surrounded by systems which are open, and are constantly being fed with energy. Numerous examples of such actively driven systems include self-assembly in biological systems, reaction-diffusion process in chemical and ecological sciences, thermal-convective phenomena in fluid dynamics, geophysical and atmospheric sciences, fracture propagation in material sciences to name a few~\cite{cross1993pattern,huber2018emergence,kuramoto1987statistical,zhang1993deterministic,chatterjee2017aging,chatterjee2016energy,georgiev2016road,chilla2012new,jaeger2010far}. The unifying theme across all of the above examples, from nanoscale to macroscale, is the staggering complexity that emerges spontaneously. As should be the case, equilibrium thermodynamics becomes insufficient in explaining the underlying dynamics anymore. Typically, far-from-equilibrium thermodynamics is treated as a natural extension of equilibrium thermodynamics. Such an approach is based on the local equilibrium hypothesis, according to which a system can be viewed as collection of subsystems where the rules of equilibrium thermodynamics hold true~\cite{kolmogorov1941local,vilar2001thermodynamics}. However, in reality a simple theoretical Carnot engine ($C$), that exchanges heat between two reservoirs maintained at different temperatures and generates work, becomes incredibly difficult to visualize in practice, see Figure 1a~\cite{casas2003temperature,garcia2008thermodynamics}. Even in order to maintain the heat baths at a constant temperature, a steady heat influx is mandatory. Thus, a practical Carnot engine ($C^\prime$) no longer remains as efficient as a theoretical Carnot engine, and its efficiency is now expressed as a function of steady-state non-equilibrium temperature of the baths and subsequent far-from-equilibrium correction, see Figure 1b.\\

\noindent The problem that we are faced with is two-fold: the absence of definition of the thermodynamic variables in a far-from-equilibrium scenario, and to be able to fit in the ideas of emergence of order and complexity into a theoretical framework. While a system at equilibrium is completely random, a system when driven out-of-equilibrium is extremely sensitive to the magnitude of the driving perturbation. For example, when water is heated over a flame, it takes some time to create a sufficient thermal gradient. Once the gradient is established a convection current is set up that drives the hotter molecules to the top and colder molecules to the bottom, in cycle. This onset of convection is denoted by the critical value of the dimensionless constant, the Rayleigh number ($Ra$). On increasing the gradient further, the convective motion becomes chaotic and turbulence sets in, which is marked by very high values of Ra ($\sim 10^9$ for a vertical surface). For a driven thermal convective system like the Rayleigh-B{\'e}nard convection, where both inertia and viscous drag play a crucial role, the region between these opposite ends of the spectrum allows for numerous possibilities in the system to let order emerge naturally~\cite{cross1993pattern,koschmieder1993benard,pandey2018turbulent}. 

\section{Methodology}
\subsection*{Rayleigh-B{\'e}nard Convection}
\begin{figure}[t]
\centering
\includegraphics[scale=0.7]{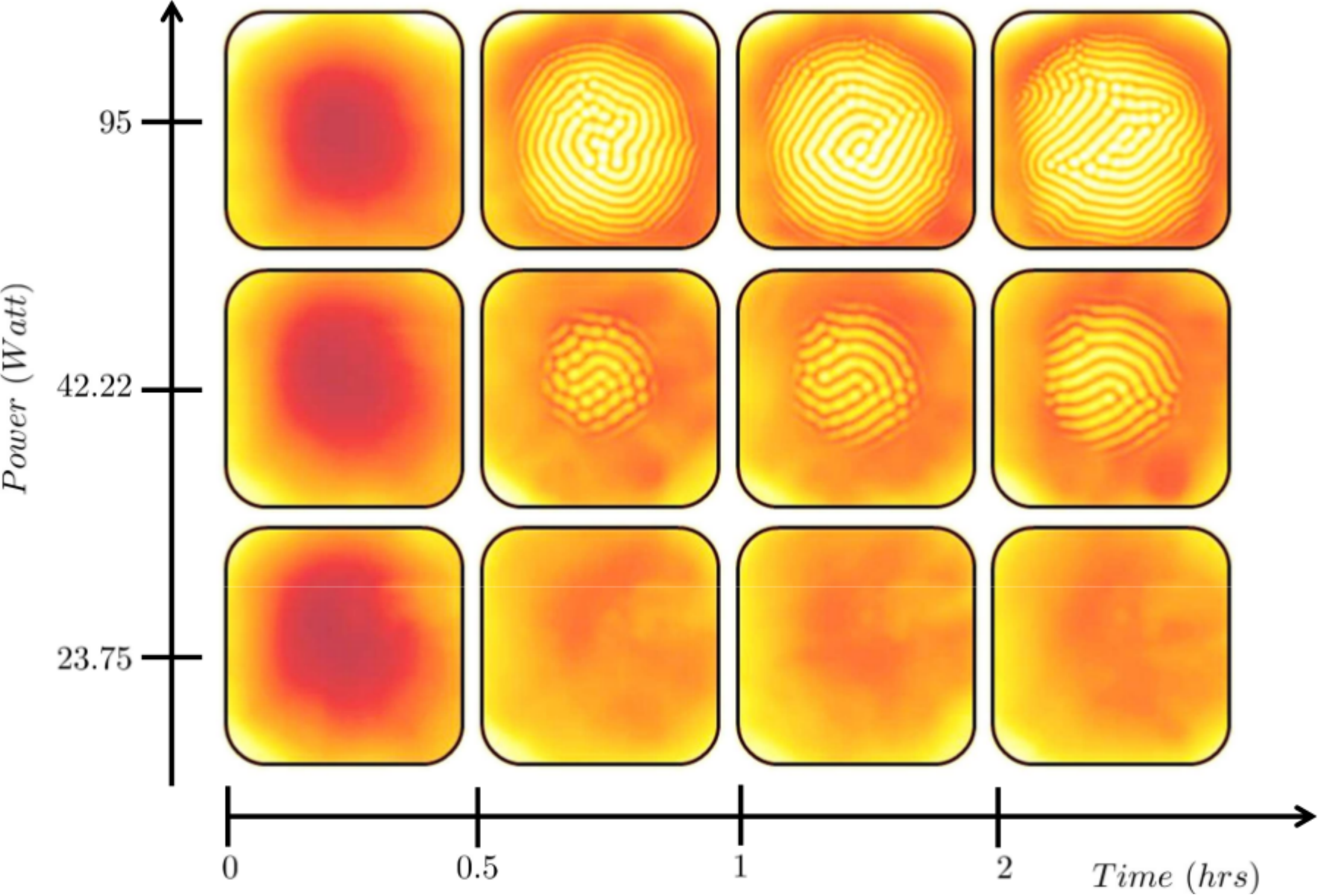}
\caption{Figure shows the thermal images of the Rayleigh-B{\'e}nard convection at different instances in time for different driving power. The right most column is the steady-state images taken after two hours with Rayleigh numbers, $Ra = 831, 2670,$ and $3464$ (bottom to top). Note that the color scale is independent to each image~\cite{chatterjee2018non,*yadati2018detailed,*chatterjeeccs17}.}
\end{figure}

\noindent The Rayleigh-B{\'e}nard convection is an excellent and perhaps the oldest prototypical model to explore the emergence of order when driven far-from-equilibrium~\cite{cross1993pattern,chilla2012new,koschmieder1993benard,pandey2018turbulent}. A thin layer of Silicone oil with a kinematic viscosity ($\nu$) of $150$ $cSt$ is heated in a Copper pan by an electric heater attached to its base. A thermocouple attached to the base of the Copper pan records the bottom temperature, $T_{bottom}$. The vertical temperature gradient (in $+z$ direction) initiates a convective motion in the fluid, and under the competing forces of buoyancy and viscosity, a thermal instability is generated. These thermal instabilities appear as patterns of various length-scales when imaged using an Infrared camera. The camera, which is calibrated by the base thermocouple temperature, $T_{bottom}$ of the empty Copper pan, is used to record the temperature of the top layer of the oil-film, $T_{top}$. The system, once fed with constant power is observed for two hours after which it reaches a steady-state.\\

\section{Discussion}

\noindent In Figure 2, we present the thermal images of the Rayleigh-B{\'e}nard convection recorded at different instants in time (along $x$-axis) as a function of increasing power (along $y$-axis). Each pixel in the thermal image corresponds to a specific temperature, $T_j$ that can be obtained by linearly extrapolating a thermal scale (white being the hottest). The color scheme in the figure thus corresponds to the respective heat maps hence generated. In the bottom most panel there is no emergent pattern as the thermal gradient between the layer thickness is not sufficiently enough to sustain thermal convection. However, comparing the top and the middle panels we can clearly see that the onset of pattern is faster at a higher power. Higher power allows for higher temperature gradient. The Rayleigh number, which is crucial to predict the onset of convection is directly proportional to the temperature difference between the layer thickness ($T_{bottom} - T_{top}$) and varies as the cube of the film thickness ($l_z$). The critical value of the Rayleigh number for the onset of convection is $1708$~\cite{koschmieder1993benard}. In the top panel, the critical Rayleigh number is achieved much faster and earlier than in the middle panel. Whereas, in the bottom panel the Rayleigh number is always less than the critical value (see Figure 2 caption for the values of the Rayleigh number at steady-state). 

\subsection*{Coexsisting Local Equilibrium States, and Free-energy Bifurcation}

\begin{figure}[t]
\centering
\includegraphics[scale=0.7]{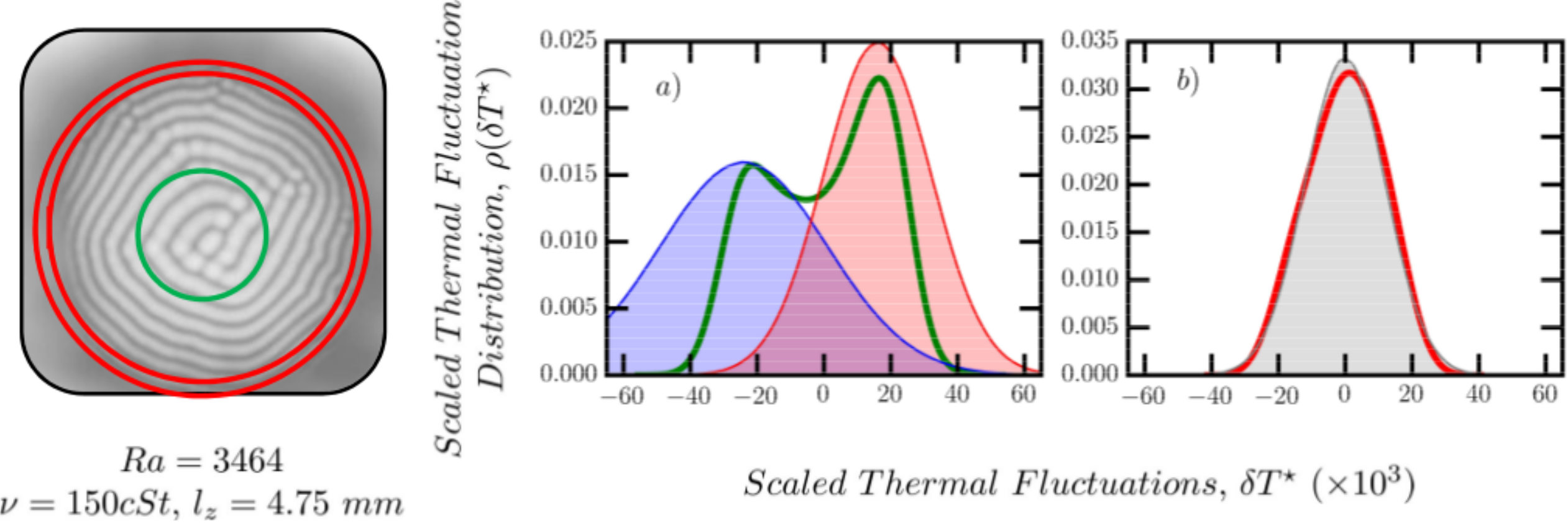}
\caption{Figures show scaled thermal fluctuation distribution for a steady-state thermal image taken at $Ra = 3464$ for (a) a circular region (dark green) and for (b) an annular region (red). The data is denoted by the line plot, the Gaussian fits are denoted by the shaded regions in the plot~\cite{chatterjee2018non,*yadati2018detailed,*chatterjeeccs17}.}
\end{figure}

\noindent In Figures 3a and 3b, we show the scaled thermal fluctuations in the far-from-equilibrium steady-state Rayleigh-B{\'e}nard convection. Thermal fluctuation at each point is defined as the variation of each pixels' temperature from the mean bulk temperature of the image. The scaled thermal fluctuation is the ratio of the thermal fluctuation to the mean bulk temperature of the system, $T^\star = \Big(\frac{T_j-<T>}{<T>}\Big)$. We can clearly observe the difference in the shapes of the fluctuation distribution plots $(\rho(T^\star))$ for the two distinct regions of interest. In Figure 3a, we select a middle region to carry out the statistics. This region captures the convection patterns, thus takes into account the presence of both the upward and downward plumes. While in Figure 3b, we select a region that is devoid of any convection cell, thus containing only thermally random currents. 

\section{Implications}

\noindent The statistical interpretations of the distributions is striking. In the region devoid of convection cells, the fluctuations are found to be random and hence a Gaussian distribution perfectly overlaps the underlying fluctuation distribution curve. However, in the region that includes the convection cells, the underlying distribution function shows a bimodality~\cite{niemela2000turbulent,grossmann2004fluctuations,chatterjee2018non}. We also point out in the figure that the slope of the distribution on the either sides is too steep for two distinct Gaussians to to be able to replicate the observed bimodality. This result points toward an interesting observation about the statistical interpretation of the system. Although, macroscopically the system is far-from-equilibrium, but microscopically, we should expect an equilibrium like behavior, as the fluxes in the system are completely balanced (steady-state). And as expected, while we do observe equilibrium-like behavior at one location, we also observe non-equilibrium fluctuations in the same system at a different spatial location. Since, the non-equilibrium fluctuation can not be replicated by two overlapping independent Gaussians, we can claim that the system, as a whole cannot be interpreted as just a linear superposition of local equilibrium states.\\

\noindent Further, we consider the interpretation of temperature for a system when driven far-from-equilibrium. Temperature is one of the principal equilibrium-thermodynamic variables that can be used as a measure of the mean energy of the particles of a matter. However, the non-equilibrium interpretations of the temperature is still unclear. Temperature in a far-from-equilibrium system varies both temporally and spatially. As we see from Figure 3a and 3b, bimodality in thermal fluctuations imply the presence of two local equilibrium temperatures in the system~\cite{chatterjee2018non,*yadati2018detailed,*chatterjeeccs17}. Since, experimental observations are made at steady-state, temperature remains constant as a function of time. But as the system has frozen into a state where multiple equilibrium points can coexist, the temperature can now be interpreted as a thermodynamic variable to represent these coexisting equilibrium points in a phase space. The trajectory which connects these local equilibrium points can then be interpreted as the energy landscape on which these local equilibrium points thrive.\\

\noindent Furthermore, the emergence of these spatial gradients on the far-from-equilibrium thermodynamic landscape due to convection implies a bifurcation in the free-energy of the system. Had the total heat transferred through the system been purely conductive, there would have been a single uniform temperature. But, convection on the other hand creates bifurcation. Macroscopically, this can be related to the phase coupling of a series of practical Carnot engines ($C^\prime$) that generates work for the convection to survive while exchanging energy between these local equilibrium points. 

\section{Conclusion}

\noindent Classical thermodynamics, one of the oldest and most successful areas of physics, based solely upon observation and deduction describes equilibrium phenomena beautifully. The vast majority of interesting phenomena and rich intricate complexities of the world around us however arise from conditions far-from-equilibrium~\cite{lieb1998guide,cross1993pattern,huber2018emergence,kuramoto1987statistical,zhang1993deterministic,chatterjee2017aging,chatterjee2016energy,georgiev2016road,chatterjee2012action,chatterjee2016thermodynamics}. Equilibrium thermodynamics can only weakly approximate far-from-equilibrium processes. Approximating far-from-equilibrium processes is often not a good idea as a systems’ response to the perturbative effects is not linear, and non-linearity possesses the natural risk of driving the system into the regime of deterministic chaos. However, in theory those phenomena in which microscopic time-scales are much faster than the macroscopic time-scales (and vice-versa), can be approximated as equilibrium processes~\cite{kolmogorov1941local,vilar2001thermodynamics}. As emergent complexity and pattern formation occur between meso and macro-scale, we set up a tabletop experiment to study far-from-equilibrium behavior. In our study we look at the far-from-equilibrium behavior in a Rayleigh-B{\'e}nard convection from a first-principles perspective. While our approach has provided us several insightful results, it has also left us with numerous open-ended questions. Through this article we try to provide some perspectives on questions like, how to interpret the system macroscopically as local order emerges when it is being driven far-from-equilibrium, or how to microscopically interpret the statistics of the distribution of states in an environment consisting of coexisting equilibrium states. Although answers to these questions are speculative at this moment, nevertheless it is in the context of our study that has led us to ask such questions in the first place. While we search for explanations, we also believe at the same time that no single theory would be probably be able to answer all of these questions due to the sheer amount of complexity present in the world around us~\cite{lieb1998guide,cross1993pattern,georgiev2016road,chatterjee2012action,chatterjee2016thermodynamics,chatterjee2018non,*yadati2018detailed,*chatterjeeccs17,kadanoff2001turbulent}. 

\section{Acknowledgments}

\noindent The authors are indebted to their collaborator Georgi Y. Georgiev (Assumption College) and the contributions of Yash Yadati and Sean McGrath. The authors are thankful for the support of the Department of Physics at Worcester Polytechnic Institute.

\end{document}